\documentclass[
reprint,
superscriptaddress,
amsmath,amssymb, aps,
prl
]{revtex4-1}

\usepackage{graphicx}
\usepackage{dcolumn}
\usepackage{bm}
\usepackage{hyperref}

\newcommand{\rom}[1]{\uppercase\expandafter{\romannumeral #1\relax}}

\newcommand{\sq}{\mathbf{S}_{\mathbf{q}}}
\newcommand{\nk}{n(\mathbf{k})}
\newcommand{\nkzero}{n(\mathbf{k}_{0})}
\newcommand{\nnss}{\langle \mathbf{S}_{i} \cdot \mathbf{S}_{j}\rangle}
\newcommand{\metal}{Metallic }
\newcommand{\tbi}{TBI }
\newcommand{\qsl}{QSL }
\newcommand{\ksl}{KAFM }

\begin{document}
	
	\title{Metal-insulator transition and intermediate phases in the kagome lattice Hubbard model}
	
	\author{Rong-Yang Sun}
	\affiliation{Kavli Institute for Theoretical Sciences, University of Chinese Academy of Sciences, Beijing 100190, China}
	
	\author{Zheng Zhu}
	\email{zhuzheng@ucas.ac.cn}
	\affiliation{Kavli Institute for Theoretical Sciences, University of Chinese Academy of Sciences, Beijing 100190, China}
	\affiliation{CAS Center for Excellence in Topological Quantum Computation, University of Chinese Academy of Sciences, Beijing, 100190, China}
	
	\date{\today}
	
	\begin{abstract}
		Motivated by the recent discovery of metallic kagome lattice materials, AV$_{3}$Sb$_{5}$ (A=K, Rb, Cs), we investigate the ground state of the half-filled kagome lattice Hubbard model by employing the density-matrix renormalization group (DMRG) method. We identify a metal-insulator transition around $U \thicksim U_{c1}$ and four distinct phases as a function of $U/t$ on narrower cylinders, including a metallic phase at $U < U_{c1}$, two insulating intermediate phases: a translational symmetry breaking phase at $U_{c1} < U < U_{c2}$ and a quantum spin liquid phase at $U_{c2} < U < U_{c3}$, and the kagome antiferromagnetic phase at $U>U_{c3}$. We confirm that the translational symmetry breaking phase is robust for wider cylinders, while the quantum spin liquid phase is smoothly connected to the kagome antiferromagnetic phase with increasing the system width. Moreover, our numerical observations indicate a continuous metal-insulator transition at $U_{c1}$, whose nature is consistent with Slater's transition scenario. The magnetic phase transition between two insulating intermediate phases at $U_{c2}$ is first order. Our findings may provide insights into exotic kagome lattice materials.
	\end{abstract}
	
	\maketitle

	\emph{Introduction.---}The recent discovery of metallic kagome materials, AV$_{3}$Sb$_{5}$ series (A=K, Rb, Cs) \cite{ortiz2019new, ortiz2020cs, jiang2020discovery, kenney2021absence, yin2021superconductivity, yu2021concurrence, zhao2021nodal, ortiz2021superconductivity, liang2021three, chen2021roton, li2021observation, chen2021highly} and T$_{m}$X$_{n}$ series (T=Mn, Fe, Co; X=Sn, Ge; $m:n$ = 3:1, 3:2, 1:1) \cite{fenner2009non, nakatsuji2015large, liu2018giant, kang2020dirac}, which exhibit either unconventional superconductivity or non-trivial topological behavior, have attracted a new round of attention on the physics emerged from kagome systems which host non-trivial electron structures (Dirac cone, van Hove singularity, and flat band) in the non-interacting limit. Previous studies mainly focus on the magnetism of the kagome Mott insulators such as the herbertsmithite ZnCu$_{3}$(OH)$_{6}$Cl$_{2}$ \cite{helton2007spin, norman2016colloquium, khuntia2020gapless} and Zn-barlowite Cu$_{3}$Zn(OH)$_{6}$FBr \cite{han2014barlowite, fu2020dynamic}, which are regarded as the most promising systems to realize quantum spin liquid (QSL) \cite{zhou2017quantum}. Theoretically, the simplest starting point to study all of the above materials is the single-band kagome-lattice Hubbard model (KHM) \cite{hubbard1963electron}; however, a systematical investigation is still a much-needed endeavor.
	
	\begin{figure}[ht]
		\centering
		\includegraphics[width=\linewidth]{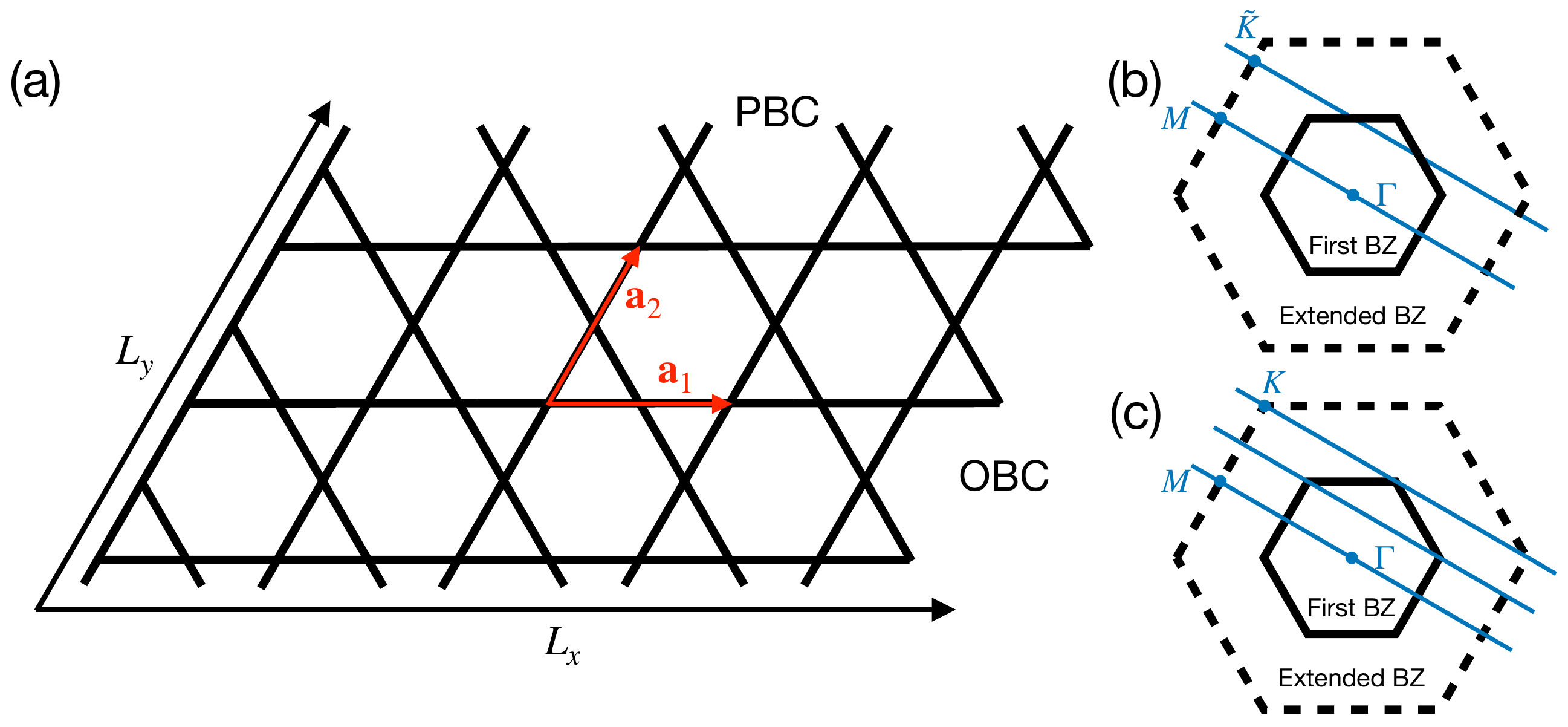}
		\caption{\label{fig:system-schematic}
			(a) Real space lattice geometry of a kagome cylinder. $\mathbf{a}_{1}$ and $\mathbf{a}_{2}$ are two primary vectors. $L_{x}$ and $L_{y}$ indicate the number of unit cells along $\mathbf{a}_{1}$ and $\mathbf{a}_{2}$ direction, respectively. The figure shows a cylinder with $L_{y} = 3$ (YC6 geometry, see \cite{kolley2015phase}) and $L_{x} = 5$. (b, c) The first and the extended Brillouin zone (BZ) of the kagome lattice. The blue cuts indicate the accessible momentum points of the finite cylinders, and the blue dots highlight the accessible (quasi-)high-symmetric points for $L_{y} = 2$ (b) and $L_{y} = 3$(c).
		}
	\end{figure}
	\begin{figure}[ht]
		\centering
		\includegraphics[width=0.9\linewidth]{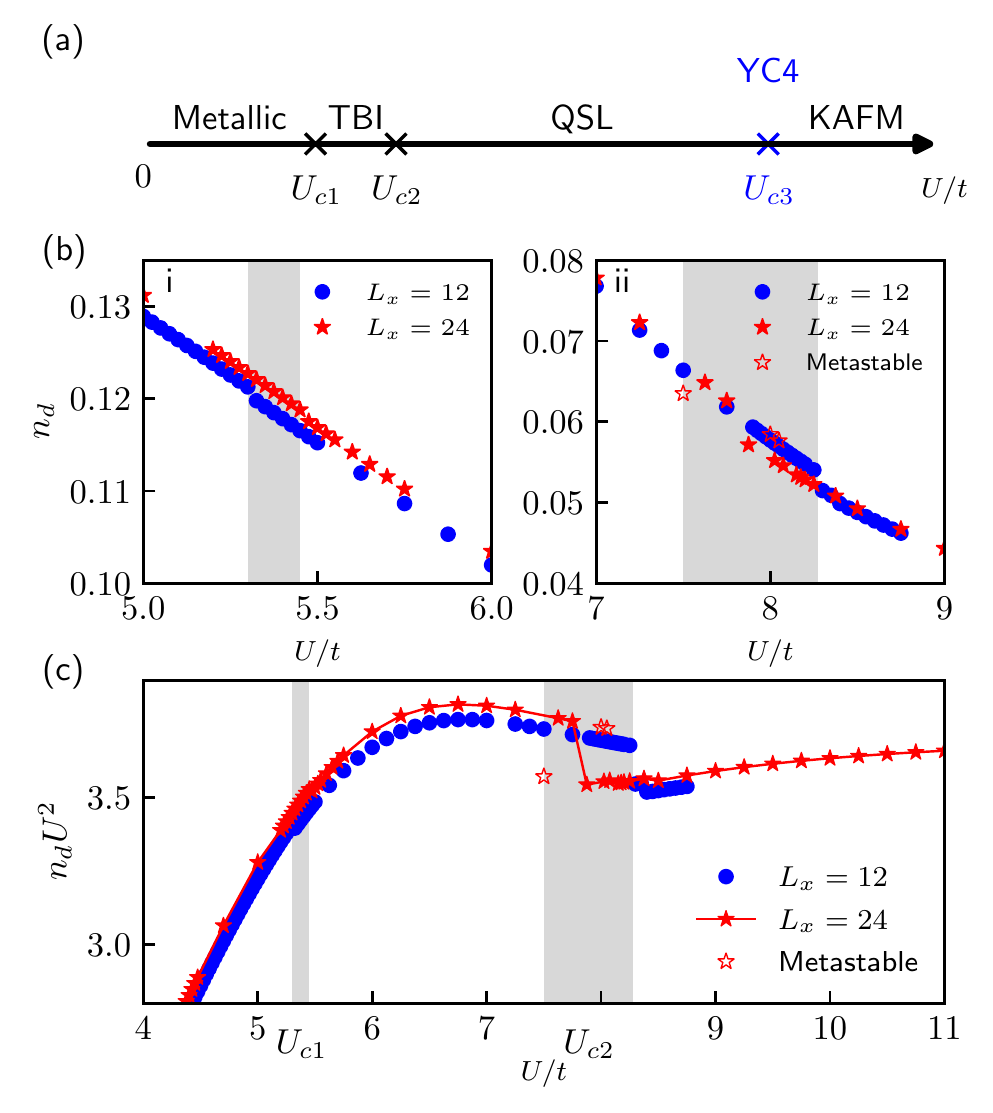}
		\caption{\label{fig:nd-vs-U}
			(a) The schematic phase diagram of the half-filled kagome lattice Hubbard model, where \tbi donates translational symmetry broken insulator, \qsl denotes quantum spin liquid, and \ksl denotes kagome antiferromagnetism. $U_{c1}$, $U_{c2}$ and $U_{c3}$ indicate the approximate phase transitions. $U_{c3}/t \sim 19.5$ only exists in YC4 cylinders. (b) Double occupancy $n_{d}$ as a function of $U/t$ near $U_{c1}/t \approx 5.4$ (i) and $U_{c2}/t \approx 7.9$ (ii) for YC4 cylinders. (c) $n_{d}U^{2}$ as a function of $U/t$ for YC4 cylinders. Gray shades indicate the possible phase boundaries. Empty red stars in (b ii) and (c) highlight these data obtained from DMRG calculations which converge to a metastable state with a slight higher energy. For more details about the hysteresis loop, see the Supplemental Material \cite{sm}.
		}
	\end{figure}
	
	The nature of the correlated phases in the KHM is controlled by the electronic correlation and electron density. The electronic correlation is represented by the ratio between the on-site Coulomb repulsion $U$ and the hopping amplitude $t$. As the research about the KHM is in the rudimentary stage, we restrict the electron density to be half-filling. In the $U/t\rightarrow \infty$ limit, the KHM reduces to the antiferromagnetic Heisenberg model (KAFM). The strong geometric frustration leads to a QSL ground state and is considered as a reliable approximation to the frustrated kagome magnets mentioned above, though the nature of the QSL is controversial \cite{ran2007projected, jiang2008density, yan2011spin, depenbrock2012nature, messio2012kagome, iqbal2013gapless, mei2017gapped, he2017signatures, liao2017gapless, chen2018thermodynamics}. With reducing the ratio of $U/t$, the charge fluctuation arises, and much less knowledge has been accumulated due to the great challenge to solve it. The interplay between the charge and spin degrees of freedom may give rise to exotic phases in the vicinity of metal-insulator transition (MIT) \cite{imada1998metal}, such as the half-filled triangular lattice Hubbard model\cite{schafer2020tracking, shirakawa2017ground, szasz2020chiral}. However, to the best of our knowledge, the report of the intermediate phase in half-filled KHM is still lacking. Furthermore, the correlation driven MIT in the KHM sounds its own importance, and its physical nature is debated. The dynamical mean-field theory (DMFT) and the quantum Monte Carlo (QMC) calculations report a first-order MIT at $U_{c}/t \sim 7~\text{to}~9$ \cite{ohashi2006mott, kaufmann2020correlations,udagawa2010chirality,bulut2005magnetic} consistent with the Brinkman-Rice picture of MIT \cite{brinkman1970application}. In contrast, the variational cluster approximation (VCA) identifies a second-order MIT at $U_{c}/t \sim 4~\text{to}~5$ with the deformation of the Fermi surface favors a valence-bond (VB) state \cite{higa2016bond} supporting Slater's approach to the MIT \cite{slater1951magnetic}. Motivated by the above experimental and theoretical progress, in this Letter, we adopt the unbiased density-matrix renormalization group (DMRG) method to systematically study the evolution of the ground state with the ratio of $U/t$ in the half-filled KHM.
	
	\emph{Model and method.---}We consider the half-filled Hubbard model on a kagome lattice described by
	\begin{equation}
		H = -t\sum_{\langle ij \rangle \sigma} (c^{\dagger}_{i\sigma} c_{j\sigma} +
		\mathrm{H.c.}) + U\sum_{i} n_{i\uparrow}n_{i\downarrow}~,
	\end{equation}
	where $c_{i\sigma}$ is the annihilation operator of the electron with spin $\sigma$ ($\sigma = \uparrow \text{or} \downarrow$) at site $i$ and $n_{i\sigma} = c^{\dagger}_{i\sigma}c_{i\sigma}$ is the on-site electron density operator. The summation of $\langle ij \rangle$ traverses all the nearest-neighbor (NN) sites on the lattice. In this work, we focus on $U>0; t>0$ region and examine the ground states of the KHM from the non-interacting limit to the strongly-correlated region by DMRG\cite{white1992density, white1993density}.
	
	We choose the cylinder geometry, as illustrated in Fig.~\ref{fig:system-schematic}~(a), the periodic (open) boundary condition is used along the $\mathbf{a}_{2}$ ($\mathbf{a}_{1}$) direction.  There are $L_{x}$ and $L_{y}$ unit cells along the $\mathbf{a}_{1}$ and $\mathbf{a}_{2}$ direction, respectively, and thus the total number of sites $N = 3 \times L_x \times L_y$. Due to the  large local dimension of the Hubbard model and the exponential increase of the DMRG computational cost with $L_{y}$, we only consider $L_{y} = 2$ (YC4) and $L_{y} = 3$ (YC6) cylinders. $L_x$ is pushed up to 36 (12) for YC4 (YC6). In our DMRG calculations, we keep the bond dimension up to 20,000 to reduce the maximum truncation error to the order or less than $10^{-9}~(10^{-4})$ for YC4 (YC6) cylinders.% The high-performance matrix product state algorithms library GraceQ/MPS2 \cite{graceq} is used to perform the simulations.

	\begin{figure}[t]
		\centering
		\includegraphics[width=0.9\linewidth]{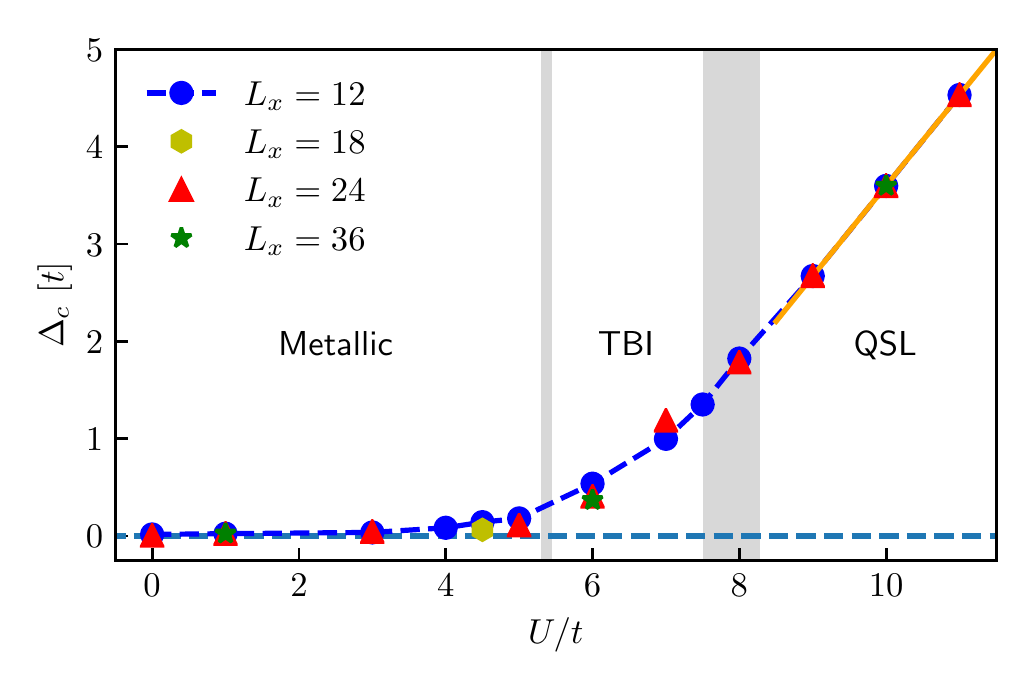}
		\caption{\label{fig:charge-gap}
			The charge gap $\Delta_{c}$ as a function of $U/t$ for the YC4 cylinder with different $L_{x}$. The gray shades indicate possible phase boundaries obtained in Figs.~\ref{fig:nd-vs-U}. The orange line in the \qsl phase is obtained by a linear fitting of $L_{x} = 24$ data to guide the eyes.
		}
	\end{figure}
	\begin{figure*}[ht]
		\centering
		\includegraphics[width=0.7\linewidth]{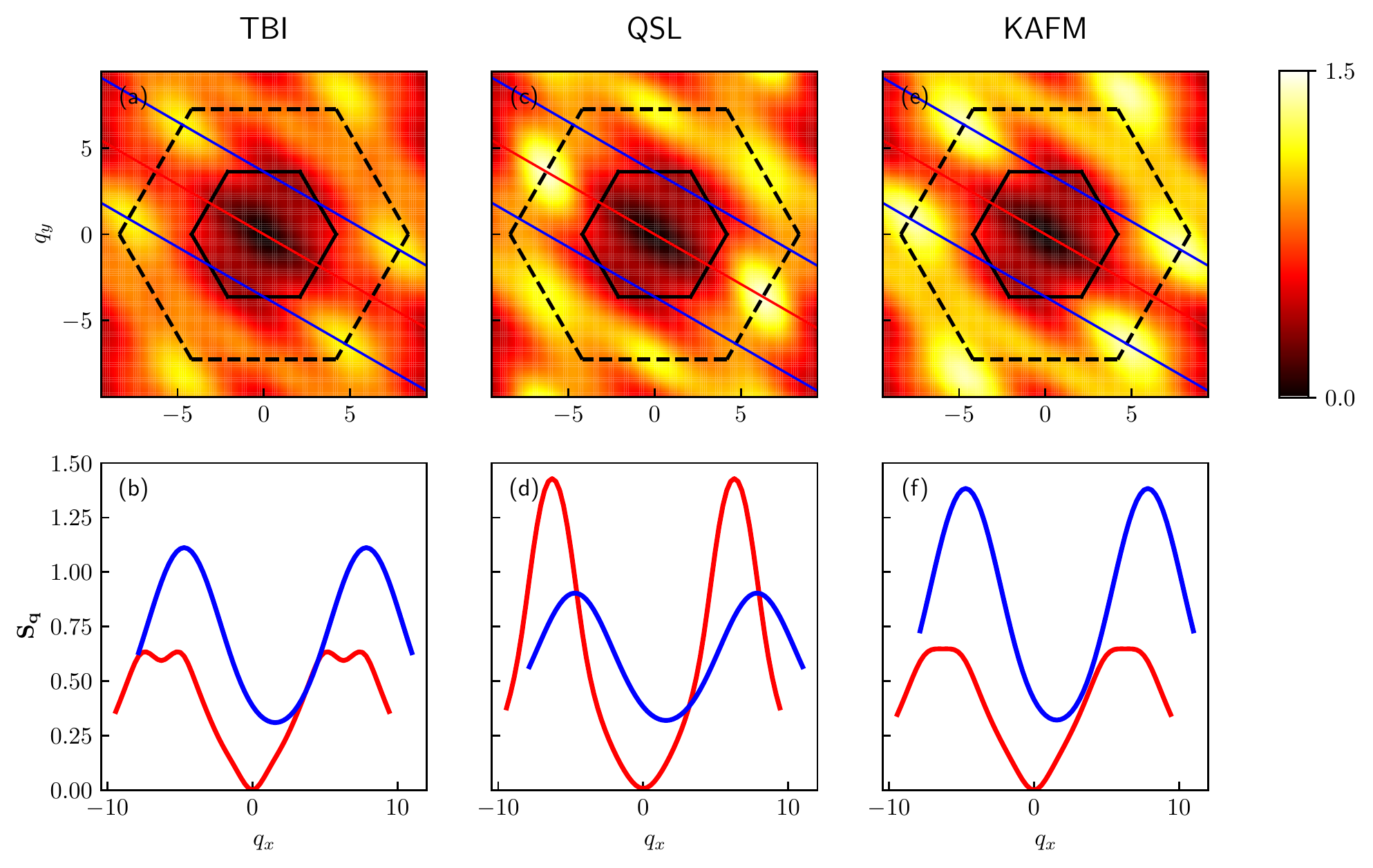}
		\caption{\label{fig:insuls-sq-yc4}
			The contour plots of the spin structure factor $\sq$(upper row) and the line cuts of $\sq$ (lower row) in the three insulating phases for YC4 cylinders with $L_{x} = 24$ [(a, b) in the \tbi phase for $U/t = 7$, (c, d) in the \qsl phase for $U/t = 10$, and (e, f) in the \ksl phase for $U/t = 20$]. The plots in the lower row illustrates the cuts of $\sq$ along the lines with the same color in the corresponding contour plot.
		}
	\end{figure*}
	\emph{Phase diagram.---}We first present the ground-state phase diagram of the half-filled KHM as a function of $U/t$ in Fig.~\ref{fig:nd-vs-U}~(a). For YC4 cylinders, we identify four distinct phases separated by three phase transition points at $U_{c1}/t \approx 5.4$, $U_{c2}/t \approx 7.9$ and $U_{c3}/t \thicksim 19.5$. The  first two phase transitions are identified by the discontinuity or kink of the site averaged double occupancy, $n_{d} = \sum_{i}\langle n_{i\uparrow} n_{i\downarrow}\rangle/N$. Figs.~\ref{fig:nd-vs-U}~(b i) and (b ii) show $n_{d}$ as a function of $U/t$ near these two phase boundaries for YC4 cylinders. At $U_{c1}$, the position of the discontinuity or kink in $n_d$  slightly shifts with $L_{x}$, but the value of $n_d$  is almost unchanged [see Fig.~\ref{fig:nd-vs-U}~(b i)], which indicates a robust phase transition in the thermodynamic limit. Since $n_{d} = \partial \epsilon_{0} / \partial U$, where $\epsilon_{0}$ is the ground-state energy per site, the kink of $n_{d}$ indicates a continuous or weak first-order phase transition at $U_{c1}$. In the vicinity of $U_{c2}$, in addition to the explicit discontinuity, we also observe its hysteresis behavior on longer cylinders with $L_{x} = 24$ [see Fig.~\ref{fig:nd-vs-U}~(b ii)], which further confirms the first-order nature of the phase transition. We would like to point out that our findings of $U_{c2}$ and its hysteresis behavior are consistent with the previous DMFT study \cite{ohashi2006mott}. $U_{c3}$ is identified by the distinct spatial patterns of the  NN spin-spin correlations $\nnss$ at its two sides (see below).  Here we point out that we only have clearly identified $U_{c3}$ on YC4 cylinders, while it is absent on YC6 cylinders, which we will discuss more below. Moreover, we also illustrate $n_{d}U^{2}$ as a function of $U/t$ across the first three phases for YC4 cylinders in Fig.~\ref{fig:nd-vs-U}~(c) to impart a panorama of the ground-state phase diagram.
	
	\emph{Evolution of charge gap with coupling strength.---}Now we explore the properties of these distinct phases. We first focus on the charge channel by probing the spin-neutral charge excitation gap, which is defined as
	\begin{eqnarray}
		\Delta_{c} = && \frac{1}{2} [ E_{0}(N_{\uparrow}+1, N_{\downarrow}+1) + E_{0}(N_{\uparrow}-1, N_{\downarrow}-1)\nonumber\\
		&&- 2E_{0}(N_{\uparrow}, N_{\downarrow})]~,
	\end{eqnarray}
	where $E_{0}(N_{\uparrow}, N_{\downarrow})$ denotes the ground state energy of a system with $N_{\uparrow}$ spin-up electrons and $N_{\downarrow}$ spin-down electrons, here $N_{\uparrow}=N_{\downarrow}=N/2$ at half filling. Figure~\ref{fig:charge-gap} shows the $\Delta_{c}$ as a function of $U/t$ for YC4 cylinders with different $L_{x}$. At the week coupling side $U<U_{c1}$, $\Delta_{c}$ approaches to zero with increasing $L_{x}$, indicating gapless charge excitation, although a tiny charge gap induced by the one-dimensionality of the system geometry cannot be ruled out. However, in the moderate coupling region with $U_{c1}<U<U_{c2}$, $\Delta_{c}$ saturates to a finite value, demonstrating finite charge gap. The opening of the charge gap at $U_{c1}$ characterizes the MIT. We also notice that the scaling behavior of $\Delta_{c}$ with $U/t$ is slightly away from linearity in the translational symmetry broken insulator (TBI) phase. Also, $\Delta_{c}$ obtained by DMRG is consistent with previous VCA results \cite{higa2016bond} in the weak and moderate coupling region. At $U>U_{c2}$, $\Delta_{c}$ is linearly proportional to $U/t$ and insensitive to $L_{x}$, in consistent with a Mott insulator with the charge gap determined by the on-site Coulomb interaction. The nature of these phases also can  be confirmed from the momentum distribution function (see the Supplementary Material \cite{sm}).
	
\begin{figure}[ht]
		\centering
		\includegraphics[width=\linewidth]{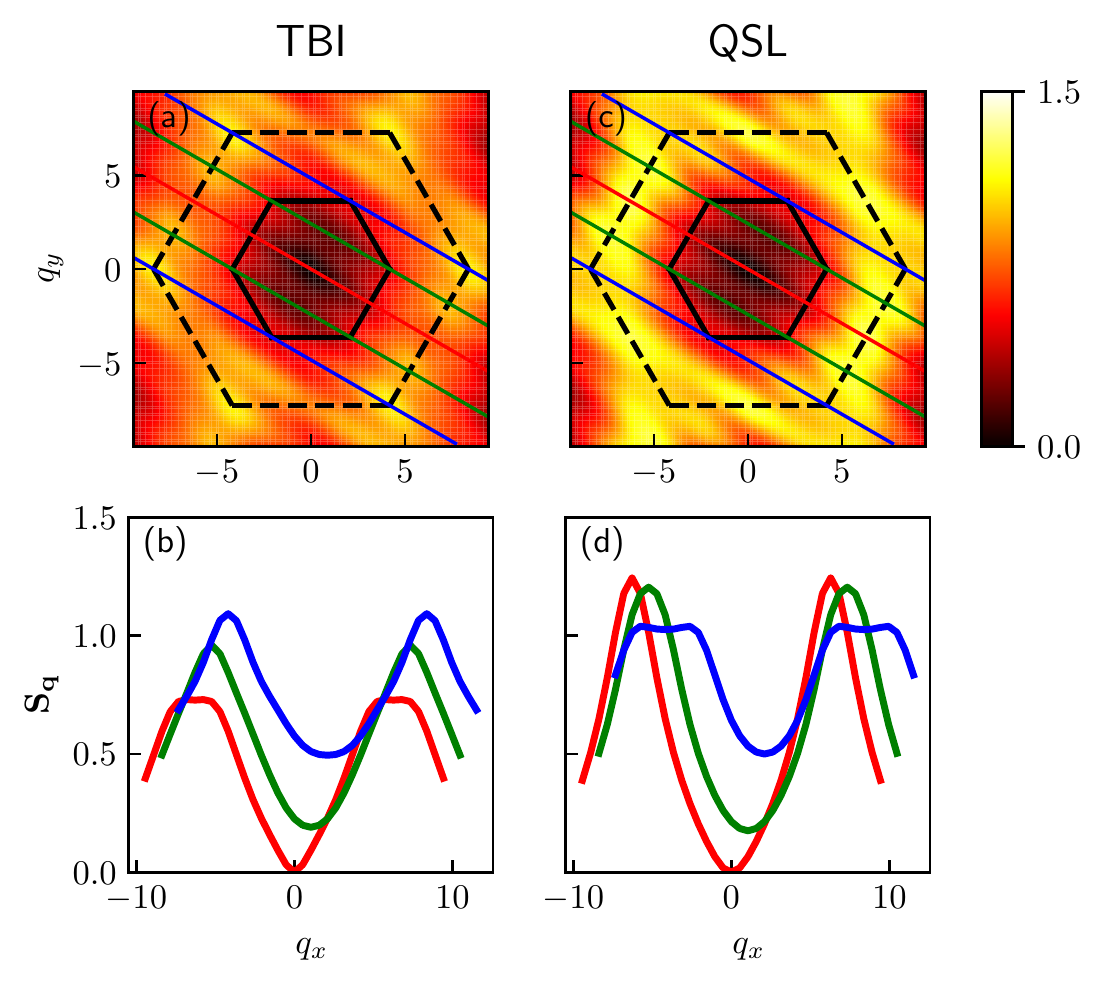}
		\caption{\label{fig:insuls-sq-yc6}
			The contour plots of the spin structure factor $\sq$ (upper row) and the line cuts  of $\sq$(lower row) in the two insulating phases for YC6 cylinders with $L_{x} = 12$ [(a, b) in the \tbi phase for $U/t = 7$ and (c, d) in the \qsl phase for $U/t = 20$]. Other legends to this figure, refer to Figs.~\ref{fig:insuls-sq-yc4}.
		}
	\end{figure}

	\emph{Spin structure factors and spatial distribution of the NN spin-spin correlations in the insulating phases.---}In the charge channel, we have identified finite charge gap at $U>U_{c1}$ [see Fig.~\ref{fig:charge-gap} for YC4 cylinders, for YC6 cylinder, we also identify $\Delta_{c} \approx 0.59 t$ for $U/t = 7$ \cite{sm}]. To further examine the nature of these insulating phases, we probe the spin channel. A straightforward way is to calculate the spin structure factor,
	\begin{equation}
		\label{eq:sq}
		\sq = \frac{1}{N} \sum_{ij}\langle \mathbf{S}_{i}\cdot\mathbf{S}_{j}\rangle  e^{i\mathbf{q}(\mathbf{r}_{i} - \mathbf{r}_{j})}~,
	\end{equation}
	where $\mathbf{S}_{i}$ is the $SU(2)$ spin operator at site $i$, and the spatial distribution of the NN spin-spin correlations $\nnss$.
	Figures~\ref{fig:insuls-sq-yc4} show the contour plots of $\sq$ and the corresponding cuts of $\sq$ through the highly symmetric points for YC4 cylinders, while Figs.~\ref{fig:insuls-sq-yc6} exhibit the same quantities in the two intermediate phases for YC6 cylinders.  The line cuts are illustrated by the lines with the same color in the corresponding contour plots.
	
	\begin{figure}[ht]
		\centering
		\includegraphics[width=0.9\linewidth]{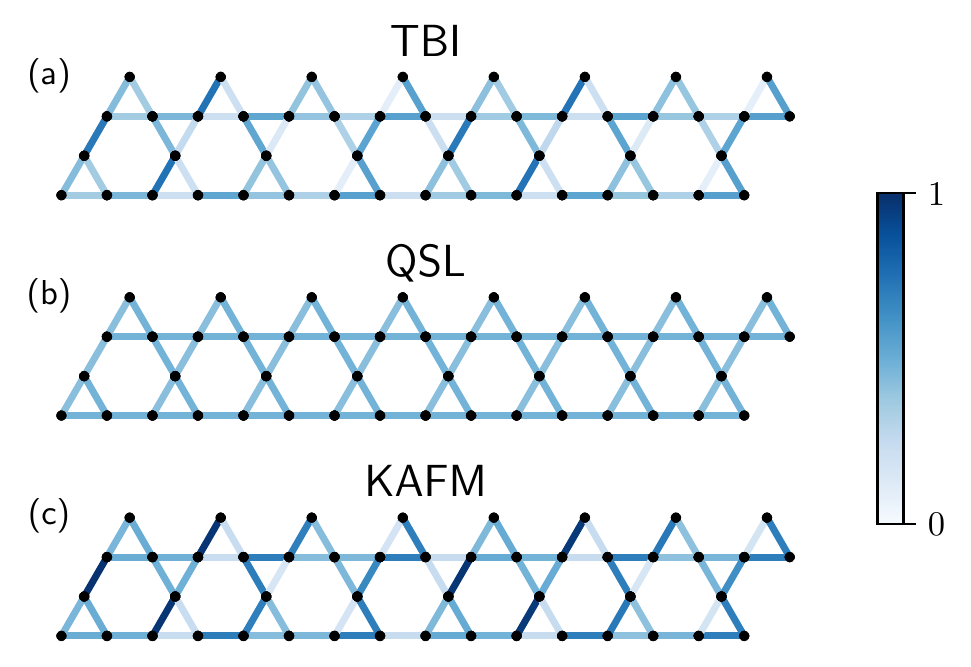}
		\caption{\label{fig:insuls-nnss-yc4}
			The normalized absolute values of NN spin-spin correlation $\nnss$ in the three insulating phases for YC4 cylinders with $L_{x} = 24$ [(a) in the \tbi phase for $U/t = 7$, (b) in the \qsl phase for $U/t = 10$, and (c) in the \ksl phase for $U/t = 20$]. We only illustrate the segment including the central 8 unit cells along $\mathbf{a}_{1}$ direction to reduce the boundary effects.
		}
	\end{figure}
	\begin{figure}[ht]
		\centering
		\includegraphics[width=0.9\linewidth]{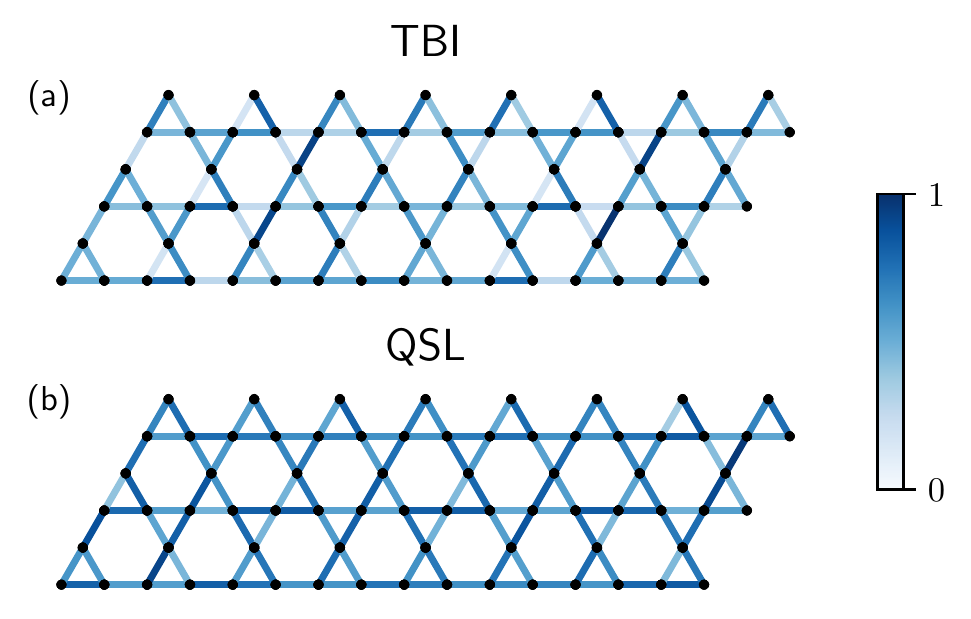}
		\caption{\label{fig:insuls-nnss-yc6}
			The normalized absolute values of NN spin-spin correlation $\nnss$ in the two insulating phases for YC6 cylinders with $L_{x} = 12$ [(a) in the \tbi phase for $U/t = 7$ and (b) in the \qsl phase for $U/t = 20$]. We only illustrate the segment including the central 8 unit cells along $\mathbf{a}_{1}$ direction to reduce the boundary effects.
		}
	\end{figure}
	
	In the \tbi phase, $\sq$ is enhanced at $\tilde{\textbf{K}}$ points [see Figs.~\ref{fig:insuls-sq-yc4}~(a, b) for YC4], while its intensity is slightly reduced on YC6 cylinder at ${\textbf{K}}$ [see Figs.~\ref{fig:insuls-sq-yc6}~(a, b)], which suggests a disordered spin background with strong commensurate spin-density-wave (SDW) fluctuations instead of SDW orders towards 2D. Meanwhile, at $U_{c1}<U<U_{c2}$, we observe a nonuniform distribution of $\nnss$ with a 4 unit cells as period [see Fig.~\ref{fig:insuls-nnss-yc4}~(a) for YC4 and Fig.~\ref{fig:insuls-nnss-yc6}~(a) for YC6], characterizing a robust translational symmetry breaking phase.
	
	In the \qsl phase, $\sq$ shows enhancement at ${\textbf{M}}$ points for YC4 [see Figs.~\ref{fig:insuls-sq-yc4}~(c, d)], but its intensity is strongly suppressed with increasing $L_y$ and $\sq$ becomes featureless [see Figs.~\ref{fig:insuls-sq-yc6}~(c, d) for YC6], which also demonstrates the absence of magnetic orders. In particular, the spatial distribution of $\nnss$ is extremely uniform for YC4 [see Fig.~\ref{fig:insuls-nnss-yc4}~(b)] and fairly uniform for YC6 cylinders [see Fig.~\ref{fig:insuls-nnss-yc6}~(b)]. All of these features are consistent with a quantum spin liquid without spin rotational or translational symmetry breaking at $U_{c2}<U<U_{c3}$ for YC4 and at $U>U_{c2}$ for YC6 cylinders. The real space spin-spin correlations are short-ranged in both \tbi and \qsl phases (for the data, see the Supplemental Material \cite{sm}), suggesting a spin gapped nature of these two phases. We also have checked the spin chirality order by computing the chiral-chiral correlations in the \tbi and \qsl phases, which we find are vanishingly small and decays rapidly, indicating the absence of time reversal symmetry breaking, in contrast to the triangular-lattice case\cite{shirakawa2017ground, szasz2020chiral, szasz2021phase, chen2021quantum, wietek2021mott}.
	
	The distinct feature of the $\sq$ and the clear first-order phase transition at $U_{c2}$ [see Fig.~\ref{fig:nd-vs-U}~(b ii)] separate the \qsl phase from the \tbi phase. Meanwhile, we find $U_{c3}$ only exists on YC4 cylinders. $U_{c3}$ is identified by the uniform spatial pattern of $\nnss$ at $U<U_{c3}$ and non-uniform pattern of $\nnss$ at $U>U_{c3}$, the latter phase exhibits the same properties as the corresponding pure spin model, indicating the map from the Hubbard model to the Heisenberg model in the $U/t\rightarrow\infty$ limit \cite{sm}. The existence of $U_{c3}$ on YC4 might be induced by the translational symmetry breaking of the ground state when putting the kagome Heisenberg model on YC4 cylinders, which finally becomes a spin liquid with increasing system width and can be smoothly connected to the \qsl phase at $U>U_{c2}$. So the QSL phase potentially shares similarities with the ground state of the kagome Heisenberg model in the 2D limit.
	
	\emph{Conclusions.---}In this work, we study the phase diagram of the half-filled kagome lattice Hubbard model as a function of $U/t$ by the unbiased DMRG method. For YC4 cylinders, we identify a metal-insulator transition at $U_{c1}$ and two nonmagnetic insulating phases at moderate coupling strength $U_{c1}<U<U_{c3}$: a translational symmetry breaking phase at $U_{c1}<U<U_{c2}$ and a quantum spin liquid phase at $U_{c2}<U<U_{c3}$. The phase at $U>U_{c3}$ is equivalent to the Heisenberg limit. For YC6 cylinders, we find the QSL phase at $U>U_{c2}$ can be smoothly connected to the Heisenberg limit while the intermediate \tbi phase keeps stable. Our numerical calculations support a possible continuous MIT at $U_{c1}$ with a Slater type transition scenario and a clear first-order magnetic transition separating two intermediate phases at $U_{c2}$. Both the metal-insulator transition and the symmetry-breaking intermediate phase present robustness when enlarging the system width. Our findings of the half-filled KHM suggest that the KHM is a promising playground to study the interplay between correlation effect and geometry frustration in a quantum many-body system. The MIT and the intermediate phases could be further tested in real kagome materials or the ultracold fermions trapped in the optical lattices \cite{bakr2009quantum, sherson2010single, parsons2016site, cheuk2016observation}, in which the coupling strength $U/t$ is widely tunable. The findings in this work might also serve as a starting point to explore the correlated phases after doping charge, which have been studied in the weakly-correlated limit \cite{kiesel2012sublattice, kiesel2013unconventional, denner2021analysis, wu2021nature}, to provide insights into the geometry effect on the doped Mott physics.
	
	\begin{acknowledgments}
		We would like to thank Yi Zhou, Yang Qi, D. N. Sheng, Gang Su, Zheng-Yu Weng, Shou-Shu Gong, and Yongqing Cai for the valuable discussions. We acknowledge the computational resources supplied by the Kavli Institute for Theoretical Sciences (KITS) at the University of Chinese Academy of Sciences (UCAS). ZZ was supported by the National Natural Science Foundation of China (Grant  No. 12074375), the Fundamental Research Funds for the Central Universities, the start-up funding of KITS at UCAS, and the Strategic Priority Research Program of CAS (No. XDB33000000). The numerical simulations in this work are based on the GraceQ project \cite{graceq}.
	\end{acknowledgments}
	

%% SM
\clearpage
\newpage
\begin{center}
	\textbf{\large Supplemental Material}
\end{center}
%%%%%%%%%% Prefix a "S" to all equations, figures, tables and reset the counter %%%%%%%%%%
\setcounter{equation}{0}
\setcounter{figure}{0}
\setcounter{table}{0}
\setcounter{section}{0}
\setcounter{page}{1}
\makeatletter
\renewcommand{\theequation}{S\arabic{equation}}
\renewcommand{\thefigure}{S\arabic{figure}}
\renewcommand{\bibnumfmt}[1]{[S#1]}
\renewcommand{\citenumfont}[1]{S#1}
%%%%%%%%%% Prefix a "S" to all equations, figures, tables and reset the counter %%%%%%%%%%
This Supplemental Material includes more numerical details: the evolution of charge momentum distribution with coupling strength, the Spatial distribution of NN spin-spin correlation for narrow KAFM cylinders, the $n_{d}$ hysteresis loop near $U_{c2}$ obtained from adiabatic evolution, the spin-spin correlations in the \tbi and \qsl phases, and the finite-size scaling of the charge gap for the YC6 cylinders in the \tbi phase.

\section{More numerical details}
\subsection{Evolution of charge momentum distribution with coupling strength}
\begin{figure}
	\centering
	\includegraphics[width=\linewidth]{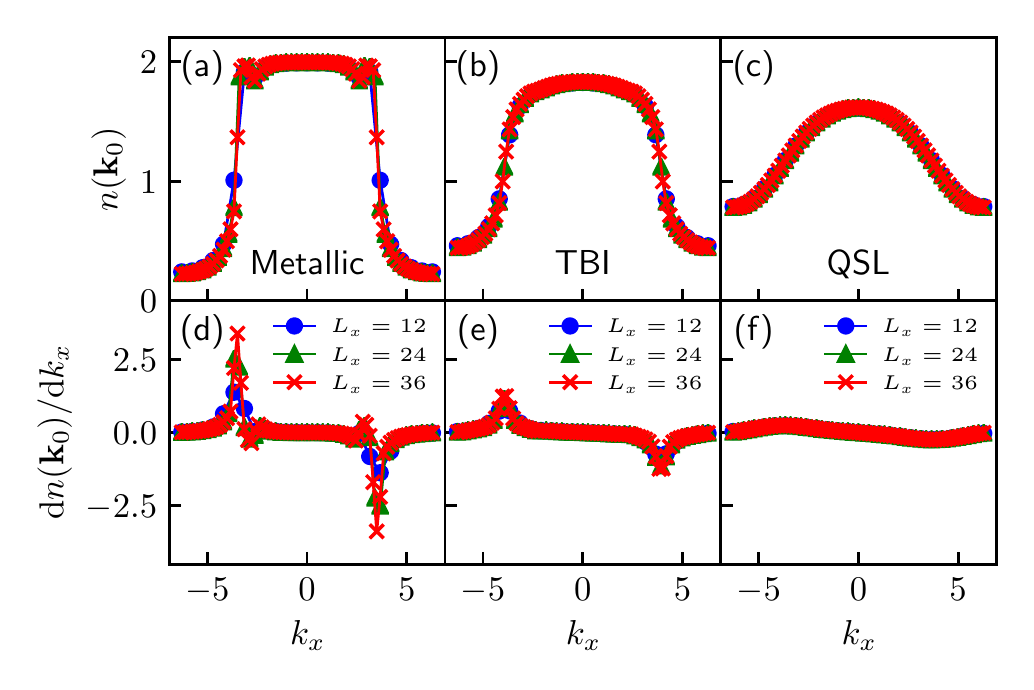}
	\caption{\label{fig:fs-evo}
		The $\nk$ distribution (a-c) and its first-order derivative (d-f) along the momentum cut which is along the $\Gamma-M$ line ($\mathbf{k}_{0}$) for YC4 cylinders, where $U/t = 1$ in the \metal phase (a, d), $U/t = 6$ in the \tbi (b, e), and $U/t = 10$ in the \qsl phase (c, f).
	}
\end{figure}

We also examine the nature of different phases by measuring the electron momentum distribution, which is defined by
\begin{equation}
	\label{eq:nk}
	\nk = \frac{1}{N}\sum_{ij\sigma}\langle c^{\dagger}_{i\sigma} c_{j\sigma}\rangle e^{i\mathbf{k}(\mathbf{r}_{i} - \mathbf{r}_{j})}~,
\end{equation}
where $i$ and $j$ traverse all lattice sites. Figures~\ref{fig:fs-evo} show the typical examples of the line cuts of $\nk$ through $\Gamma-M$ line [refer to Figs.~1~(b, c)], where its momenta are labeled as $\mathbf{k}_{0}$,  and the corresponding first-order derivative $\text{d}n(\mathbf{k}_{0}) / \text{d}k_{x}$  in three distinct phases for YC4 cylinders.

At $U<U_{c1}$, $\nkzero$ distribution exhibits sudden jumps [see Fig.~\ref{fig:fs-evo}~(a)], which become even sharper with increasing $L_{x}$ as illustrated by  $\text{d}n(\mathbf{k}_{0}) / \text{d}k_{x}$ in Fig.~\ref{fig:fs-evo}~(d), suggesting a metallic phase with well-defined Fermi surfaces and finite quasi-particle weight. In the \tbi phase, the jumps of $\nkzero$ becomes smoother and saturates with the increase of $L_{x}$ [see Figs.~\ref{fig:fs-evo}~(b, e)], consistent with charge gap opening in this moderate coupling region. In the \qsl phase, $\nkzero$ tends to be flat with increasing $U/t$ and is independent on $L_{x}$ [see Figs.~\ref{fig:fs-evo}~(c, f)], similar to the usual behavior of the Mott insulators, where $\nkzero$ would be totally flat in the $U/t \rightarrow\infty$ limit.

%\bigskip

\subsection{Spatial distribution of NN spin-spin correlation for narrow KAFM cylinders}
\begin{figure}
	\centering
	\includegraphics[width=\linewidth]{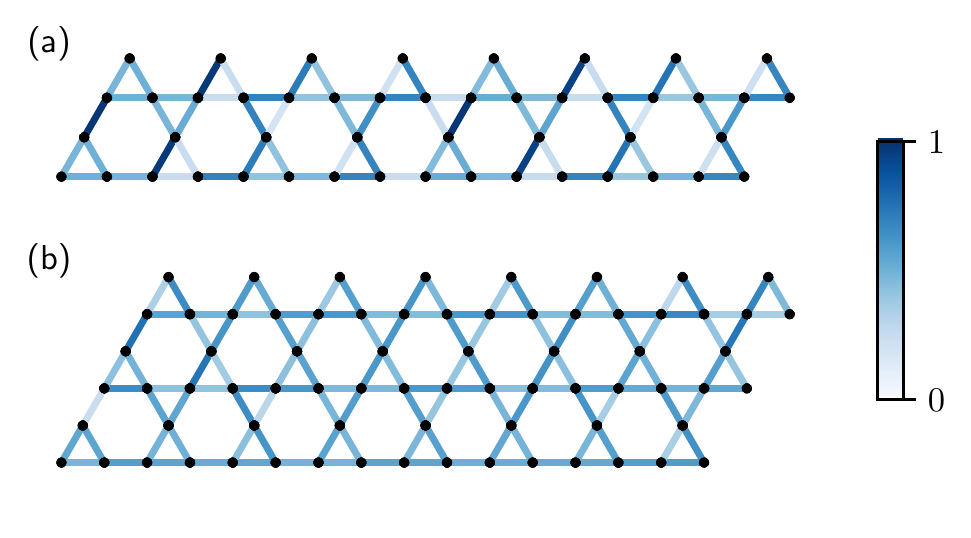}
	\caption{\label{fig:nnss-kafm}
		The spatial distribution of the normalized absolute value of NN spin-spin correlation for KAFM cylinders. (a) YC4 cylinder with $L_{x} = 24$; (b) YC6 cylinder with $L_{x} = 12$. We only illustrate the segment including the central 8 unit cells along $\mathbf{a}_{1}$ direction to reduce the boundary effects. The bond dimension of such calculation is set to $8000\sim12000$.
	}
\end{figure}

To compare with the large $U$ region of KHM, we perform DMRG calculations of pure spin KAFM model for YC4 and YC6 cylinders. For YC4 cylinder, the spatial distribution of NN spin-spin correlations [see Fig.~\ref{fig:nnss-kafm}~(a)] breaks the translational symmetry with the same VB pattern as the \tbi phase [see Fig.~6~(a)] and \ksl phase [see Fig.~6~(c)] for KHM model. For YC6 cylinder, the $\nnss$ exhibits a fairly uniform distribution [see Fig.~\ref{fig:nnss-kafm}~(b)] consisting with the \qsl phase [see Fig.~7~(b)] for YC6 KHM model.

\newpage

\subsection{$n_{d}$ hysteresis loop near $U_{c2}$}
\begin{figure}
	\centering
	\includegraphics[width=0.8\linewidth]{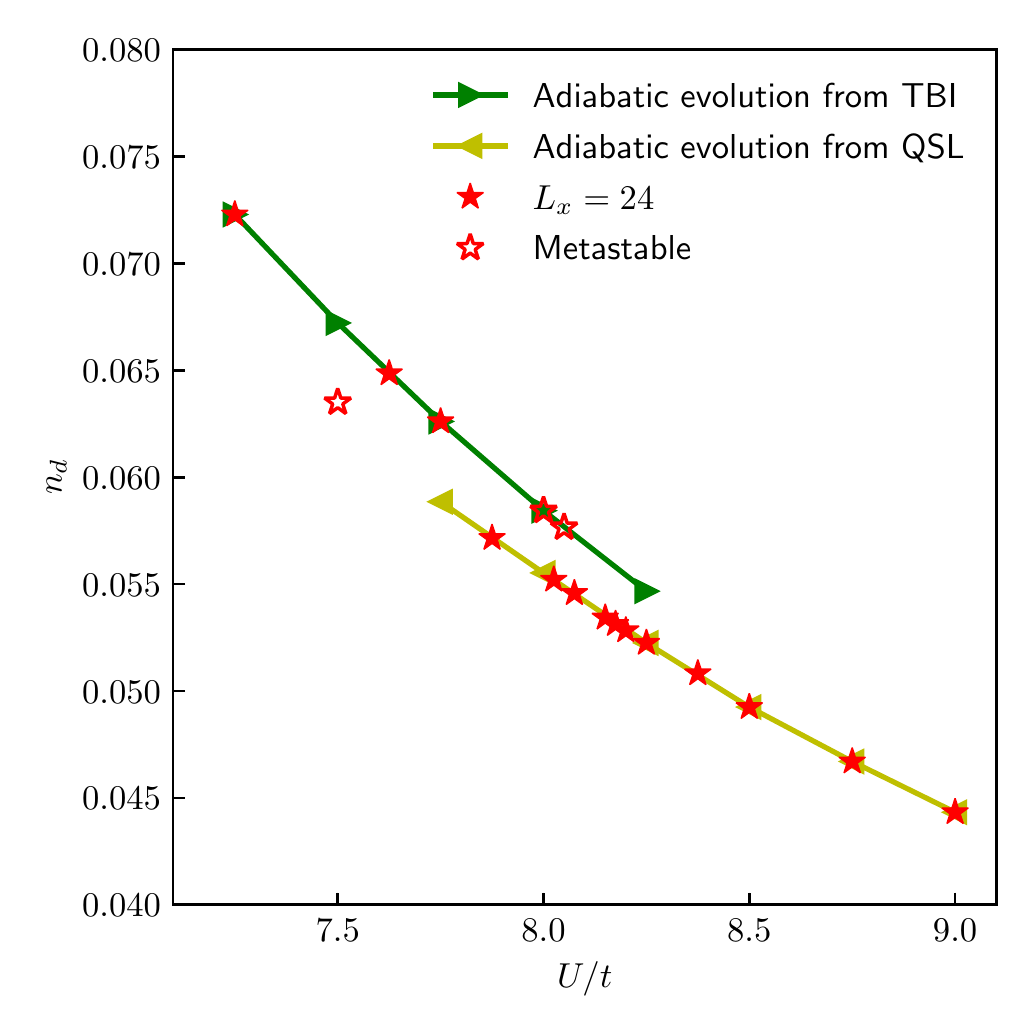}
	\caption{\label{fig:nd-near-uc2}
		The $n_{d}$ as a function of $U/t$ near the $U_{c2}$. The solid red stars show the data obtained from the simulations converged to the global minimum. The empty red stars show the data obtained from the simulations converged to the metastable state. The green (yellow) triangles show the data simulated from a converged ground state with smaller (larger) $U/t$ in the same line except for the leftmost and the rightmost ones. They are the global minimums.
	}
\end{figure}

Expect for the hysteresis loop constructed by the \emph{true} ground states and metastable states (in the main text), we can also verify the hysteresis behavior at this first-order phase transition point by examining the converged states obtained from the DMRG simulation with a particular initial wave function. To check the metastable region of the \tbi phase, we start the DMRG calculation using a well-converged state in the \tbi phase, which has a smaller $U/t$. After this simulation is converged, we adiabatically tune the $U/t$ to a larger one until the system goes into the \qsl phase, and this \tbi state becomes a metastable state. Similarly, we can also adiabatically evolve the \qsl ground state into the \tbi phase. These two adiabatic paths form a hysteresis loop illustrated in Figure~\ref{fig:nd-near-uc2} which is consistent with the hysteresis loop in the main text.

\subsection{Spin-spin correlations in the \tbi and \qsl phases}
\begin{figure}
	\centering
	\includegraphics[width=0.8\linewidth]{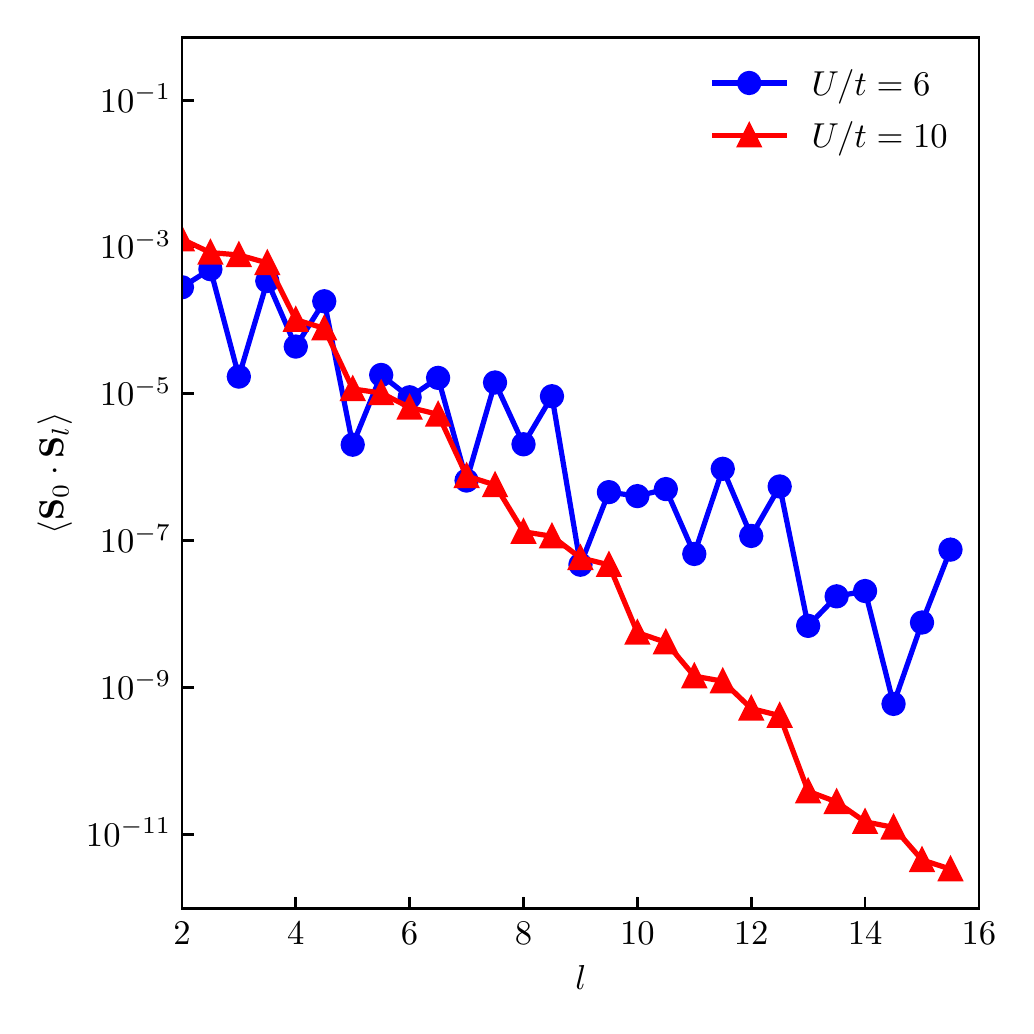}
	\caption{\label{fig:ss-insuls}
		The real space spin-spin correlation function in the \tbi ($U/t = 6$) and \qsl ($U/t=10$) phase for YC4 cylinders in a semi-logarithmic plot. The reference point is set to the 12th column, and the total system length is 36.
	}
\end{figure}

The real space spin-spin correlation function in the \tbi and \qsl phases for YC4 cylinders are presented in Figure~\ref{fig:ss-insuls}. We can see the correlations decays exponentially and are very short-ranged in both \tbi and \qsl phases, which indicate the gapped nature of the spin sector in these two phases.

\subsection{Finite-size scaling of the charge gap for the YC6 cylinders in the \tbi phase}
\begin{figure}
	\centering
	\includegraphics[width=0.9\linewidth]{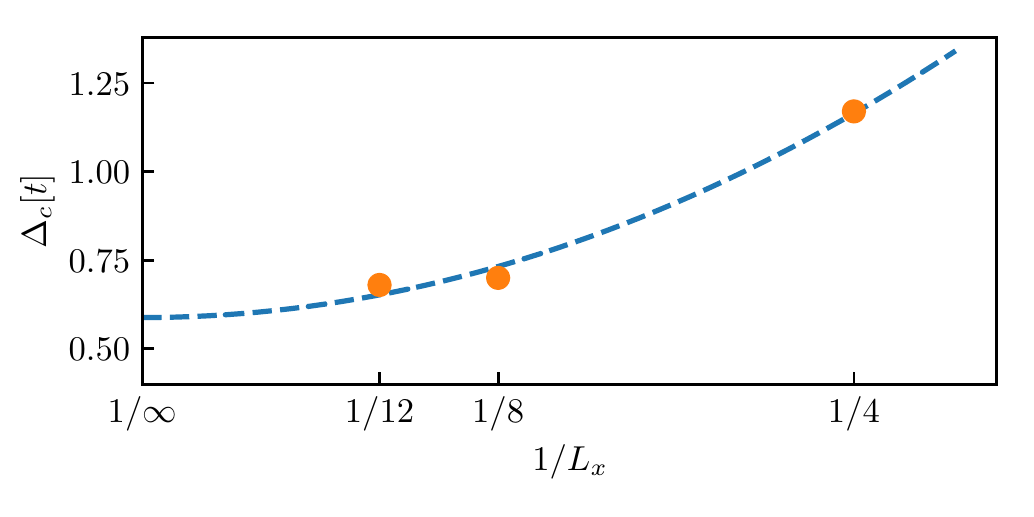}
	\caption{\label{fig:yc6-charge-gap}
		The charge gap as a function of $1/L_{x}$ in the \tbi($U/t=7$) phase for YC6 cylinder. The dashed line indicates the scaling function curve.
	}
\end{figure}

The finite-size scaling of the charge gap in the  \tbi ($U/t=7$) phase for YC6 cylinders is presented in Figure~\ref{fig:yc6-charge-gap}. The scaling function is $\Delta_{c}(L_{x}) = a \times (1/L_{x})^{2} + b$, where $a$ and $b$ are undeterminated parameters. We can see the charge gap extrapolates to $\Delta_{c} \approx 0.59 t$ in the thermodynamic limit.
\end{document}